\documentclass[12pt]{article}
\pdfoutput=1 
     
\usepackage{hyperref}    
\usepackage{amsmath}       
\usepackage{amssymb} 
\usepackage{graphicx}
\usepackage{url}

\usepackage{float,wrapfig,color} 

\allowdisplaybreaks
\def\eq#1{(\ref{#1})}
\def\s[#1\s]{\begin{align}\begin{split}#1\end{split}\end{align}}
\def\[#1\]{\begin{align}#1\end{align}}

\def\tR{{\tilde R}}
\def\tk{{\tilde k}}
\def\tS{{\tilde S}}

\def\Xint#1{\mathchoice
   {\XXint\displaystyle\textstyle{#1}}%
   {\XXint\textstyle\scriptstyle{#1}}%
   {\XXint\scriptstyle\scriptscriptstyle{#1}}%
   {\XXint\scriptscriptstyle\scriptscriptstyle{#1}}%
   \!\int}
\def\XXint#1#2#3{{\setbox0=\hbox{$#1{#2#3}{\int}$}
     \vcenter{\hbox{$#2#3$}}\kern-.5\wd0}}

\def\dashint{\Xint-}

\begin{document}

\begin{titlepage}

\title{
\hfill\parbox{4cm}{ \normalsize YITP-20-102}\\ 
\vspace{1cm} 
Symmetry enhancement in a two-logarithm matrix model \\
and the canonical tensor model}

\author{
Naoki Sasakura\footnote{sasakura@yukawa.kyoto-u.ac.jp}
\\
{\small{\it Yukawa Institute for Theoretical Physics, Kyoto University,}}
\\ {\small{\it  Kitashirakawa, Sakyo-ku, Kyoto 606-8502, Japan}}
}

\date{\today}

\maketitle

\begin{abstract}
I study a one-matrix model of a real symmetric matrix with a potential
which is a sum of two logarithmic functions and a harmonic one. 
This two-logarithm matrix model is the absolute square norm of a toy wave function
which is obtained by replacing the tensor argument of the wave function of the canonical tensor model (CTM)
with a matrix. 
I discuss a symmetry enhancement phenomenon in this matrix model and show that 
symmetries and dimensions of emergent spaces are stable only in a phase which exists exclusively
for the positive cosmological constant case in the sense of CTM.
This would imply the importance of the positivity of the cosmological constant in the emergence phenomena
in CTM.
\end{abstract}

\end{titlepage}

\section{Introduction}
\label{sec:introduction}
Quantum gravity is one of the serious fundamental problems in theoretical physics.
This problem originates from the fact that it is difficult to apply the standard quantum field theoretical
method to the quantization of general relativity\footnote{However, see for example \cite{Reuter:2019byg} 
for a sophisticated quantum field theoretical approach.}.  
Various approaches have been proposed, and some of them argue that macroscopic spacetimes 
and general relativity are emergent phenomena from collective dynamics of some fundamental 
degrees of freedom \cite{Loll:2019rdj,Rovelli:2014ssa,Surya:2019ndm,Konopka:2006hu}. 

The tensor model may be regarded as one of such approaches 
\cite{Ambjorn:1990ge,Sasakura:1990fs,Godfrey:1990dt,Gurau:2009tw}. 
It was introduced as an extension of the matrix model,  
which successfully describes the two-dimensional quantum gravity  \cite{DiFrancesco:1993cyw}, 
with the hope to extend the success to higher dimensions.
However, the tensor model does not seem to
generate macroscopic spacetimes but is rather dominated by singular objects like branched 
polymers\footnote{See \cite{Bonzom:2011zz,Gurau:2011xp} for the proof in the large $N$ limit in the colored tensor
model.}.
Therefore, it seems difficult to regard the tensor model as quantum gravity for dimensions higher than two.

The tensor model above is in the context of Euclidean simplicial quantum gravity.
In fact, simplicial quantum gravity is more successful in the Lorentzian context: 
It has been shown that the causal dynamical triangulation, the Lorentzian version, successfully generates
macroscopic spacetimes \cite{Ambjorn:2004qm}, 
while the dynamical triangulation, the Euclidean version, does not. 
Prompted by the success, the present author has formulated a tensor model in the Hamilton formalism,
which we call the canonical tensor model (CTM) \cite{Sasakura:2011sq,Sasakura:2012fb}.
CTM is  a first-class constrained system having an analogous structure as 
the Arnowitt-Deser-Misner (ADM) formalism of general relativity.
The canonical quantization of CTM is straightforward \cite{Sasakura:2013wza}, 
and the physical state condition can exactly be solved by a wave function \cite{Narain:2014cya}. 

The wave function is represented by a multiple integral of an integrand which has an argument of 
a real symmetric tensor $P_{abc}\ (a,b,c=1,\ldots,N)$ \cite{Narain:2014cya} (See \ref{app:ctm} for
a minimal introduction.). 
It has been argued in general contexts and has been shown for some simple cases 
that the wave function has peaks at Lie-group 
symmetric configurations (namely, $P_{abc}=g_{a}^{a'}g_{b}^{b'}g_{b}^{b'}P_{a'b'c'},\ g\in G$) 
for various Lie-group representations $G$ \cite{Obster:2017pdq,Obster:2017dhx}.
This phenomenon, which may be called symmetry emergence from quantum coherence, 
would be interesting from the 
perspective of spacetime emergence, since spacetimes could potentially be realized as gauge orbits of 
Lie group representations\footnote{In fact, the peaks of the wave function of CTM form a ridge
along configurations invariant under a Lorentzian Lie group $SO(n,1)$ rather than a Euclidean Lie group
\cite{Obster:2017dhx}}. 
However, we need more thorough knowledge of the phenomenon to argue for spacetime emergence,
including large $N$ limits, in which continuum spacetimes are expected to appear.

In the previous works \cite{Lionni:2019rty,Sasakura:2019hql,Obster:2020vfo,Sasakura:2020jis}, we studied the wave function in the negative cosmological constant case through a matrix model with 
non-pairwise index contractions.
However, real interesting properties of CTM are expected to appear in the positive 
cosmological constant case, since the symmetry emergence phenomenon mentioned above is 
much more evident in the positive case than the negative \cite{Obster:2017dhx}.  
Here the Monte Carlo simulations performed in the previous 
works \cite{Sasakura:2019hql,Obster:2020vfo,Sasakura:2020jis} cannot easily be applied, 
because the quantity to be computed for the positive case suffers from the notorious sign problem
of Monte Carlo simulations. 

In this paper, we consider a matrix version of the wave function of CTM by replacing the tensor argument
$P_{abc}$ with a matrix $M_{ab}$.  
This of course is not an approximation to the wave function, but its similarity makes the correspondence 
of the parameters and the interpretations between the matrix and the tensor versions possible.
An advantage of the matrix version is that it can be computed even for the positive cosmological constant case,
as we will see.
As will be explained in Section~\ref{sec:connection}, the matrix model we consider comes from 
the absolute square norm
of the matrix version of the wave function, and is given by a one-matrix model of a real symmetric  
matrix $M_{ab}\ (a,b=1,2, \cdots,N)$ with a partition function defined by
\[
Z=\int_{\mathbb{R}^{\#M}} \prod_{a,b=1\atop a\leq b}^N dM_{ab}\, e^{-S(M)},
\label{eq:partition}
\]    
where $\#M=N(N+1)/2$, namely, the number of independents components of $M_{ab}$, and 
\[
S(M):={\rm Tr} \left[ \frac{R}{2} \log(k_1+i k_2-i M) +\frac{R}{2} \log(k_1-i k_2+i M)+\alpha M^2 \right]
\label{eq:action}
\]
with positive parameters, $R,k_1,k_2,\alpha$.
The parameters have a redundancy under the rescaling of $M_{ab}$, and $\alpha=1$ may be taken
in the following sections.

A similar matrix model with two logarithmic functions has been considered  in a different context 
in \cite{Paniak:1995ef} with a difference of the last term in \eq{eq:action}.

\section{Connection to the canonical tensor model}
\label{sec:connection}
As explained in \ref{app:ctm}, \eq{eq:varphiai} gives the wave function corresponding to the exactly solved physical state \cite{Narain:2014cya} of CTM mentioned in Section~\ref{sec:introduction}. 
We consider an analogous wave function which is obtained by replacing $P_{abc}$ with $M_{ab}$:
\s[
&\Psi(M):=\langle M| \Psi \rangle=\varphi(M)^R,\\
&\varphi(M):=\int_{\mathbb{R}^N} \prod_{a=1}^N d\phi_a\, e^{i M_{ab} \phi_a \phi_b -(k_1+i k_2) \phi_a \phi_a},
\label{eq:wavefn}
\s]
where the repeated indices are assumed to be summed over.
Here the integration region is the whole $N$-dimensional real space, $M_{ab}$ is a real symmetric 
matrix, and $k_1,k_2,R$ are assumed to be positive\footnote{The sign of $k_2$ can always be chosen positive
by the replacement $M\rightarrow -M$.}.
The part containing $k_1,k_2$ of the integrand is an analogue to the Airy function in \eq{eq:varphiai}.
If $k_1$ dominates, the part becomes a damping function corresponding to 
the negative cosmological constant case in CTM, 
but, if $k_2$ dominates, the part becomes oscillatory corresponding
to the positive cosmological constant case. 
As explained in Section~\ref{sec:introduction}, since we are mainly interested in the positive
cosmological constant case, our main focus is on the case with finite $k_2$ and small $k_1$. 
More precisely, $\tilde k_1=k_1/\sqrt{N}\ll 1$ (which will appear later) is implicitly assumed
throughout this paper.  

Let us consider the following observable for the state, 
\[
\langle \Psi | e^{-\alpha \hat M^2} | \Psi \rangle = \int_{{\mathbb R}^{\# M}} \prod_{a,b=1 \atop a\leq b}^N dM_{ab} \left | \Psi(M) \right |^{2} \, e^{-\alpha M^2}, 
\label{eq:obs}
\]
where $\alpha$ is a positive parameter, $M^2:=M_{ab}M_{ab}$, 
and the integration is over the whole $\# M$-dimensional real space.
By performing the gaussian integration over $\phi_a$ in \eq{eq:wavefn} and putting the result into \eq{eq:obs}, 
we obtain
\[
\langle \Psi | e^{-\alpha \hat M^2} | \Psi \rangle =const. \, Z,
\]
where $Z$ is given in \eq{eq:partition}, and the overall constant is irrelevant.

It would be instructive to cast the same system into a different expression. 
Let us assume $R$ is an integer.
Then the $R$-th power of the wave function \eq{eq:wavefn}
can be replaced by introducing $R$ replicas of $\phi_a$: 
\[
\Psi(M)=
\varphi(M)^{R}=\int_{\mathbb{R}^{NR}} \prod_{a,l=1}^{N,R} d\phi_a^l\, e^{\sum_{l=1}^{R}
i M_{ab} \phi_a^l \phi_b^l -(k_1+i k_2) \phi_a^l \phi_a^l}.
\]
Considering the same replacement for the complex conjugate $\Psi^*(M)$ with variable $\tilde \phi_a^l$, 
putting them into \eq{eq:obs}, and integrating over $M$, we obtain
\[
Z=const.\int \prod_{a,l=1}^{N,R} d\phi_a^ld\tilde \phi_a^l\ e^{-S_{\phi}},
\]  
where
\[
S_{\phi}:= \frac{1}{4 \alpha} 
\left(
\hbox{Tr}\left[\phi  \phi^t \phi \phi^t\right]+\hbox{Tr}\left[\tilde \phi \tilde \phi^t \tilde \phi \tilde \phi^t\right]-2\hbox{Tr}\left[\phi \phi^t \tilde \phi \tilde \phi^t\right]\right)+(k_1+i k_2) \hbox{Tr}\left[\phi \phi^t\right] + 
(k_1-i k_2)\hbox{Tr}\left[\tilde \phi \tilde \phi^t\right], 
\label{eq:8v}
\]
where $(\phi\phi^t)_{ab}:=\sum_{l=1}^{R} \phi_a^l\phi_b^l$.
This may be regarded as a special choice of the parameters of the 8-vertex matrix model 
presented in \cite{Kazakov:1998qw}.
This can also be regarded as a usual matrix analogue to the matrix model with 
non-pairwise index contractions which has been analyzed in \cite{Lionni:2019rty,Sasakura:2019hql,Obster:2020vfo,Sasakura:2020jis} 
in the context of CTM.

\section{Aligned Coulomb gas picture}
\label{sec:coulomb}
In this section, we give an intuitive picture of the dynamics of the matrix model 
\eq{eq:partition} by regarding it as an aligned Coulomb gas system. 
A solid treatment of the matrix model by the Schwinger-Dyson equation will be discussed in Section~\ref{sec:sd}.

The partition function of the matrix model can be rewritten by using its invariance under 
the $SO(N)$ transformation $M'=L M L^t, \ L\in SO(N)$.  
Denoting the eigenvalues of $M$ by $\lambda_a\ (a=1,2,\ldots,N)$, which are all real, one obtains
\[
Z&=const. \int_{\mathbb{R}^N} \prod_{a=1}^N d\lambda_a \prod_{a,b=1 \atop a<b}^N | \lambda_a -\lambda_b|
\ e^{-\sum_{a=1}^N S(\lambda_a)} \\
&= const. \int_{\mathbb{R}^N} \prod_{a=1}^N d\lambda_a \, e^{-{S_{Coul}(\lambda)}}
\] 
where $\prod_{a<b} | \lambda_a -\lambda_b|$ is the Jacobian for the change of variable  
from $M$ to $\lambda_a$ (and integrate over $L$), and 
\[
S_{Coul}(\lambda):=-\sum_{a,b=1 \atop a<b}^N \log |\lambda_ a-\lambda_b |+R \sum_{a=1}^N \log |\lambda_a -k_2+i k_1 |
+\alpha \sum_{a=1}^N \lambda_a^2.
\label{eq:scoulomb}
\] 

The form of $S_{Coul}(\lambda)$ in \eq{eq:scoulomb} 
shows that the eigenvalue system can be interpreted as 
a system of charged particles on a line interacting with each other by the two-dimensional 
Coulomb potentials.
More precisely, 
the first term represents that the particles of unit charges are located at $\lambda_a\ (a=1,2,\ldots,N)$ on $\mathbb{R}$ and interact with each other by the Coulomb repulsive potentials. 
The second term can be interpreted as that there exists an opposite charge $-R$ 
located at a fixed location $k_2$ interacting with the particles of unit charges by its Coulomb potential.
Here $k_1$ can be regarded as a sort of small regularization parameter to the potential,
since, as explained in Section~\ref{sec:connection}, our main interest is the case of small $k_1$
corresponding to the positive cosmological constant case in CTM. 
The third term represents a harmonic potential for all the particles.

\begin{figure}
\begin{center}
\includegraphics[width=3.5cm]{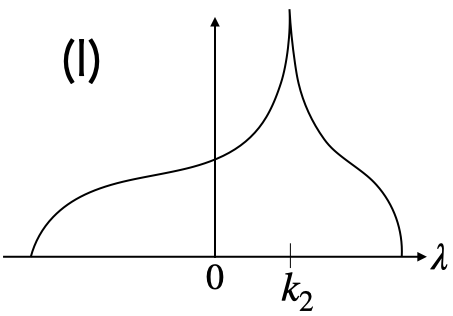}
\hfil
\includegraphics[width=3.5cm]{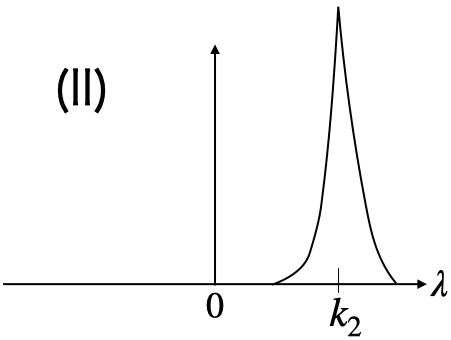}
\hfil
\includegraphics[width=3.5cm]{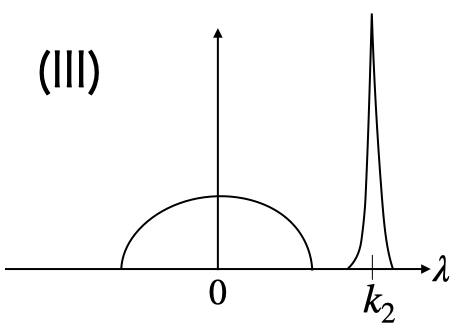}
\hfil
\includegraphics[width=3.5cm]{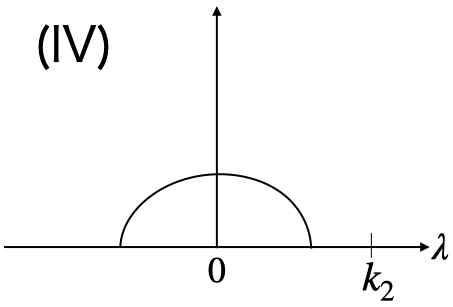}
\caption{The four possible profiles of the eigenvalue distributions for the matrix model \eq{eq:partition}.}
\label{fig:distribution}
\end{center}
\end{figure}

While the first and the third terms generate the eigenvalue distribution of the semi-circle 
law \cite{Wigner}, the second term generated by the $-R$ charge attracts the particles 
to the neighborhood of $k_2$, and part of the $-R$ charge is screened.
Therefore we can expect the following four possibilities of the eigenvalue distributions to occur:
\begin{itemize}
\item[(I)] For small $R$ and small $k_2$, there is a semi-circle-like distribution with a 
concentration (peak) around $k_2$.
\item[(II)] For large $R$, all the eigenvalues concentrate around $k_2$.
\item[(III)] For large $R$ and large $k_2$, there are two bunches of eigenvalues, one with a semi-circle-like
distribution and the other around $k_2$.
\item[(IV)] For small $R$ and large $k_2$, the eigenvalues form a semi-circle-like distribution
with no concentration around $k_2$.
\end{itemize}
The four cases are illustrated in Figure~\ref{fig:distribution}.
Here note that all the parameters $R,k_1,k_2,\alpha$ are assumed to be positive, as mentioned below \eq{eq:action}.

Next let us discuss the large $N$ limit in a similar manner as \cite{Brezin:1977sv}. 
To balance all the terms in \eq{eq:scoulomb}, the scaling with $N$ can be determined to be
\s[
R&=N \tilde R, \\
k_i&= \sqrt{N} \tilde k_i, \\
\lambda_a &= \sqrt{N} \tilde \lambda_a,
\label{eq:rescaling}
\s]
where $\tilde R$, $\tilde k_i$, and $\tilde \lambda_a$ are supposed to be kept finite in the limit.
By introducing a distribution function $\rho(\tilde \lambda)$ for the eigenvalues, the $S_{Coul}$ in \eq{eq:scoulomb} can be rewritten in the large $N$ limit as
\[
S_{cont}(\rho)=N^2 \left[- \frac{1}{2} \int_{\mathbb{R}^2} dx dy\, \rho(x) \rho(y) \log|x-y|+
\int_{\mathbb{R}} dx\, \rho(x) \left( \tilde R \log|x-\tilde k_2+i \tilde k_1|+\alpha x^2 \right) \right],
\]
where we have ignored an additional irrelevant constant, and have taken the normalization of $\rho$ as 
\[
\int_\mathbb{R} dx\, \rho(x)=1
\label{eq:normalization}
\]
for $\sum_{a=1}^N \cdots \rightarrow N \int_{\mathbb{R}} dx\, \rho(x) \cdots$.

After adding the Lagrange multiplier $\beta (\int dx \rho(x)-1)$ to 
take into account the constraint \eq{eq:normalization},
the functional derivative of $S_{cont}$ with respect to $\rho(x)$ leads to 
the stationary equation,
\[
-\int_\mathbb{R} dy\, \rho(y) \log|x-y| + \tilde R \log|x-\tilde k_2+i \tilde k_1|+\alpha x^2+\beta=0.
\label{eq:stationary}
\]
Note that this is valid only at $x$ where $\rho(x)> 0$. This is because there is an implicit constraint 
$\rho(x)\geq 0$ and the functional derivative cannot freely be taken in the invalid region.
Taking further the derivative of \eq{eq:stationary} 
with respect to $x$, it is obtained that
\[
\dashint_\mathbb{R} dy \frac{1}{x-y} \rho(y) = \frac{\tilde R(x-\tilde k_2)}{(x-\tilde k_2)^2+\tilde k_1^2} 
+ 2 \alpha x,
\label{eq:forcebalance}
\]
where the integration with a dash represents the Cauchy principal value. 
This equation will be treated by using the Schwinger-Dyson equation  
of the matrix model in the large $N$ limit in Section~\ref{sec:sd}.

\section{Analysis by the Schwinger-Dyson equation}
\label{sec:sd}
In this section we will study the matrix model \eq{eq:partition} in the large $N$ 
limit by the Schwinger-Dyson equation.
See for example \cite{DiFrancesco:1993cyw} for some details of the techniques used
in this section.

Let us define
\[
W(z):=\frac{1}{N}\left\langle {\rm Tr} \left[ \frac{1}{z-M} \right]\right\rangle,
\label{eq:defofW}
\]
where $z$ is a complex variable, and $\langle \cdot \rangle$ denotes the expectation value 
in the matrix model. 
Since we consider real symmetric $M$, the singularities of $W(z)$ can 
only be on the real axis. As we will see below, $W(z)$ has branch cuts of square roots
on the real axis, and the eigenvalue density $\rho(x)$ on real $x$ is related to $W(z)$ by
\[
\rho(x)=\frac{i}{2\pi}\left[ W(x+i\, \epsilon)-W(x-i \,\epsilon)\right]
\label{eq:rwrelation}
\] 
with $\epsilon=+0$.

Let us consider the Schwinger-Dyson equation,
\[
\int_{\mathbb{R}^{\#M} }\prod_{a,b=1 \atop a\leq b}^N dM_{ab} \,
D^M_{cd}  \left\{ \left( \frac{1}{z-M} \right)_{cd} e^{-S(M)} \right\}=0,
\label{eq:sdfirst}
\] 
where $D^M_{ab}$ denotes the partial derivative with respect to $M_{ab}$ defined by 
\[
D^M_{ab} M_{cd}=\delta_{ac}\delta_{bd}+\delta_{bc}\delta_{ad}
\]
with the symmetric property of the indices of $M_{ab}$ being taken into account.
By taking the derivatives on the lefthand side of \eq{eq:sdfirst}, we obtain
\s[
\int \prod_{a,b=1 \atop a\leq b}^N dM_{ab} &\left\{ 
\left( \frac{1}{z-M} \right)_{cc}\left( \frac{1}{z-M} \right)_{dd} + \left( \frac{1}{z-M} \right)_{cd}\left( \frac{1}{z-M} \right)_{dc}\right. 
\\
&\left. -2 \left( \frac{1}{z-M} \right)_{cd}S'(M)_{dc} \right\}e^{-S(M)}=0.
\label{eq:sdeq}
\s] 
Let us perform the large $N$ limit given in \eq{eq:rescaling}, where the last one corresponds to 
$M\rightarrow \sqrt{N} M$ (accompanied with $z\rightarrow \sqrt{N} z$).
Then, in the leading order of $N$, by assuming the factorization for the first term and ignoring the second 
term as sub-leading, we obtain an equation,
\[
W(z)^2-\frac{2}{N} \left \langle {\rm Tr} \left[\frac{\tilde S'(M)}{z-M} \right] \right\rangle=0,
\label{eq:sd2}
\]
where
\[
\tilde S'(M) =  \frac{\tR}{2(M-\tk_2 -i \tk_1)}+\frac{\tR}{2(M-\tk_2 +i \tk_1)}+ 2 \alpha M.
\]
By applying the partial fraction decomposition to the last term of \eq{eq:sd2}, it can further 
be rewritten as
\[
W(z)^2 -2 \tS'(z) W(z) +\tR\left( \frac{W(\tk_2+i \tk_1)}{z-\tk_2-i\tk_1} +\frac{W(\tk_2-i \tk_1)}{z-\tk_2+i \tk_1}\right)+ 4 \alpha=0.
\]
Therefore, the solution is given by
\[
W(z)=\tS'(z)-\sqrt{\tS'(z)^2 -\left[\tR\left( \frac{W(\tk_2+i \tk_1)}{z-\tk_2-i\tk_1} +\frac{W(\tk_2-i \tk_1)}{z-\tk_2+i \tk_1}\right)+ 4 \alpha\right]}.
\label{eq:sdsolution}
\]
Here the branch of the square root must appropriately be chosen as will be explained below. 

The solution \eq{eq:sdsolution} has a complex free parameter $W(\tk_1+i \tk_2)$ 
($W(\tk_1- i \tk_2)$ is its complex conjugate.). 
The parameter has to be tuned so that the solution 
becomes consistent with the expected properties of $W(z)$ defined in \eq{eq:defofW}: 
$W(z)$ has the asymptotic behavior $W(z)\sim 1/z$ for $z\rightarrow \infty$;  
singularities (actually cuts) are only on the real axis. 
One can check that the asymptotic behavior and the absence of the poles of $\tS'(z)$ in $W(z)$ 
are automatically satisfied by the solution \eq{eq:sdsolution} 
with an appropriate choice of the branch of the square root. 
However, it is difficult to tune $W(\tk_1+i \tk_2)$ so that all the branch cuts of \eq{eq:sdsolution} be located only on the real axis. This is because the content of the square root in \eq{eq:sdsolution}
is expressed as a sixth order polynomial function of $z$ over a common denominator, 
and the dependence of its zeros on $W(\tk_1+i \tk_2)$ is too complicated to analyze.
Therefore we rather take a different strategy to determine $W(z)$, which will be explained below.
Whether a solution by the method below corresponds to the solution \eq{eq:sdsolution} 
with a value of $W(\tk_1+i \tk_2)$
can be checked afterwards for each solution.  

From the discussions about the possible eigenvalue distributions in Section~\ref{sec:coulomb} and 
\eq{eq:rwrelation}, we expect $W(z)$ has one or two cuts on the
real axis. From the form of $\tS'(z)$, we can assume the following forms of $W(z)$.
For the cases (I), (II), or (IV) in Section~\ref{sec:coulomb}, we assume a one-cut solution, 
\[
W(z)=\tS'(z)-\left(\frac{c}{z-\tk_2-i \tk_1}+\frac{c^*}{z-\tk_2+i \tk_1}+2 \alpha\right) \sqrt{z-c_+}\sqrt{z-c_-},
\label{eq:formw1}
\]
where $c$ is generally complex,  $c_-$ and $c_+$ are real with $c_-< c_+$,
and the square roots are taken in the principal branch. From \eq{eq:rwrelation}, $\rho(x)$ is 
non-zero in the region $c_- < x <c_+$.
For the case (III),  we assume a two-cut solution,
\[
W(z)=\tS'(z)-\frac{2 \alpha (z-c)}{(z-\tk_2)^2+\tk_1^2} \sqrt{z-c_1} \sqrt{z-c_2} \sqrt{z-c_3} \sqrt{z-c_4},
\label{eq:formw2}
\]
where the parameters are assumed to be all real and satisfy $c_1< c_2 < c < c_3 < c_4$, and 
the square roots are taken in the principal branch. 
From \eq{eq:rwrelation}, the eigenvalue density $\rho(x)$ is non-zero in the two regions, 
$[c_1,c_2]$ and $[c_3,c_4]$. The reason for imposing $c_2 < c< c_3$
is that this condition is necessary for $\rho(x)$ to be positive in both the two regions.

For the expressions \eq{eq:formw1} and \eq{eq:formw2},
it is straightforward to write down the conditions for the absence of singularities except on the real axis
and the asymptotic behavior $W(z)\sim1/z$ in $z\rightarrow \infty$.
For the one-cut solution \eq{eq:formw1}, we obtain
\s[
&\tR/2-c\, \sqrt{\tk_2+i\tk_1-c_+}\,\sqrt{\tk_2+i\tk_1-c_-}=0,\\
&-\alpha ( c_++c_-) +c+c^*=0, \\
&\tR+(c+c^*)(c_++c_-)/2+\alpha (c_+-c_-)^2/4-c (\tk_2+i \tk_1)-c^*(\tk_2-i \tk_1)=1.
\label{eq:cond1}
\s]
The first condition comes from that $W(z)$ should not inherit the poles of $\tS'(z)$, 
since they are at complex values $z=k_2 \pm i k_1$.
The second and the third conditions are for the asymptotic behavior of $W(z)$ to be $\sim 1/z$.
Since the number of the conditions and that of the free parameters are the same, the solution
is uniquely determined (There may exist some discrete sets of solutions, though).

For the two-cut solution \eq{eq:formw2}, 
we first perform a replacement of the argument $z-\tk_2=y$ for the simplicity
of the expressions, and parameterize $W(z)$ as 
\[
W(y)=\tS'(y+\tk_2)-\frac{2 \alpha (y-d)}{y^2+\tk_1^2} \sqrt{y-d_1} \sqrt{y-d_2} \sqrt{y-d_3} \sqrt{y-d_4},
\]
where $d=c-\tk_2$ and so on, and $d_1< d_2 < d < d_3 <d_4$. 
We obtain
\s[
&\frac{\tR}{2}-\frac{\alpha(i \tk_1-d)}{i \tk_1} \prod_{l=1}^4 \sqrt{i \tk_1-d_l}=0, \\
&\tk_2+d+\frac{1}{2} \sum_{l=1}^4 d_l=0,\\
&\tR+2 \alpha \left(
-\frac{d}{2} \sum_{l=1}^4 d_l+\tk_1^2+\frac{1}{8}\sum_{l=1}^4 d_l^2-\frac{1}{4}\sum_{l,m=1 \atop l<m}^4 d_l d_m \right)=1.
\label{eq:cond2}
\s]
The first condition is for the absence of the poles of $\tilde S'(z)$ in $W(z)$, and the last two 
are for the asymptotic behavior. 
The conditions \eq{eq:cond2} give four real conditions in total, since the first is a complex valued condition, 
and the latter are real. 
Therefore the solution has one free parameter, since the number of the parameters in \eq{eq:formw2} is five.

The presence of one free parameter in the two-cut solution is physically understandable by the 
aligned Coulomb gas picture of Section~\ref{sec:coulomb}. 
There are two bunches of the eigenvalues, between which there exists an infinite potential barrier in the 
large $N$ limit. Therefore part of the eigenvalues can, freely to some extent, be moved 
between the two bunches without loosing the stability of the solution. 
This freedom gives one free parameter to the solution.

Yet one can fix this freedom by imposing that the two bunches have the same chemical potential \cite{Jurkiewicz:1990we}.
In other words, this condition is that there is no energy cost when moving an eigenvalue from one bunch 
to the other. This balance between the two bunches is relevant 
if $N$ is not taken strictly to the infinite.
To obtain the condition for the balance, let us take the aligned Coulomb gas picture of Section~\ref{sec:coulomb}. 
Suppose a particle of a small charge is located at $x$ between the two bunches of the particles. 
Then the force $F(x)$ acting on the particle is proportional to
\[
F(x)=W(x)-\tS'(x),
\label{eq:force}
\]  
where the first term represents the repulsive Coulomb forces coming from the particles contained in the 
bunches, and the second the forces from the negative charge and the harmonic potential. 
Then the energy cost of moving a particle from one bunch to the other 
is given by $\int_{c_2}^{c_3} dx \, F(x)$. This should vanish for the balance, 
and, by using \eq{eq:formw2}, we obtain
\s[
\int_{d_2}^{d_3} dy\, \frac{y-d}{y^2+\tk_1^2} \sqrt{y-d_1} \sqrt{y-d_2} \sqrt{d_3-y} \sqrt{d_4-y}=0.
\label{eq:chemical}
\s]  
With this additional condition, the two-cut solution is uniquely determined (There may exist some discrete 
sets of solutions, though).

Once a solution is obtained, one can compute the eigenvalue density by \eq{eq:formw1}.
We obtain
\s[
&\rho(x)=\frac{1}{\pi} \left(\frac{c}{x-\tk_2-i \tk_1}+\frac{c^*}{x-\tk_2+i \tk_1}+2 \alpha\right) \sqrt{c_+-x}\sqrt{x-c_-}
\label{eq:rho1}
\s]
in the region $[c_-,c_+]$ for the one-cut solution. 
Note that the positivity of $\rho(x)$ in the one-cut solution is not automatically satisfied,
and there actually exist solutions with negative regions of $\rho(x)$ for some parameters.
In such a case, the solutions are not correct, and other one-cut or two-cut solutions
must be taken for these parameters.
For the two-cut solution \eq{eq:formw2}, once a solution is found with $c_2<c<c_3$, 
the positivity of $\rho(x)$ is automatically satisfied and 
\s[
&\rho(x)=\frac{1}{\pi} \frac{2 \alpha |x-c|}{(x-\tk_2)^2+\tk_1^2} \sqrt{|(x-c_1)(x-c_2)(x-c_3)(x-c_4)|}
\label{eq:rho2}
\s]
for the ranges $[c_1,c_2]$ and $[c_3,c_4]$.

\section{Examples and a simple case}
\label{sec:sol}
\begin{figure}
\begin{center}
\includegraphics[width=4cm]{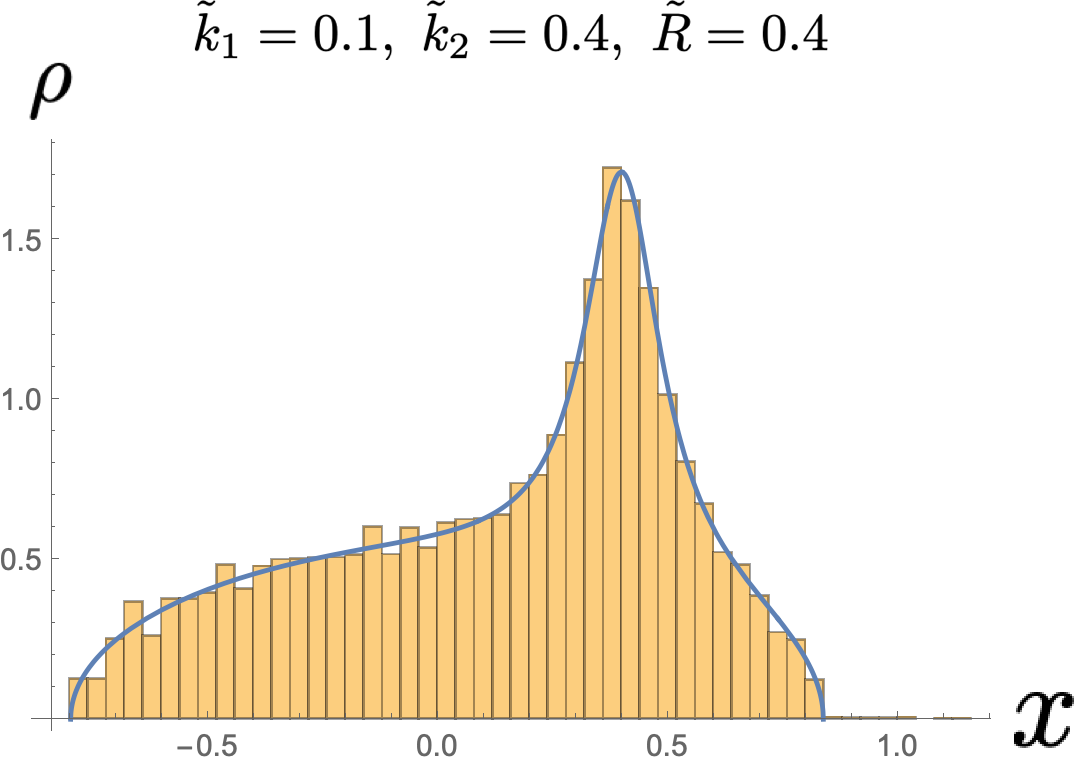}
\includegraphics[width=4cm]{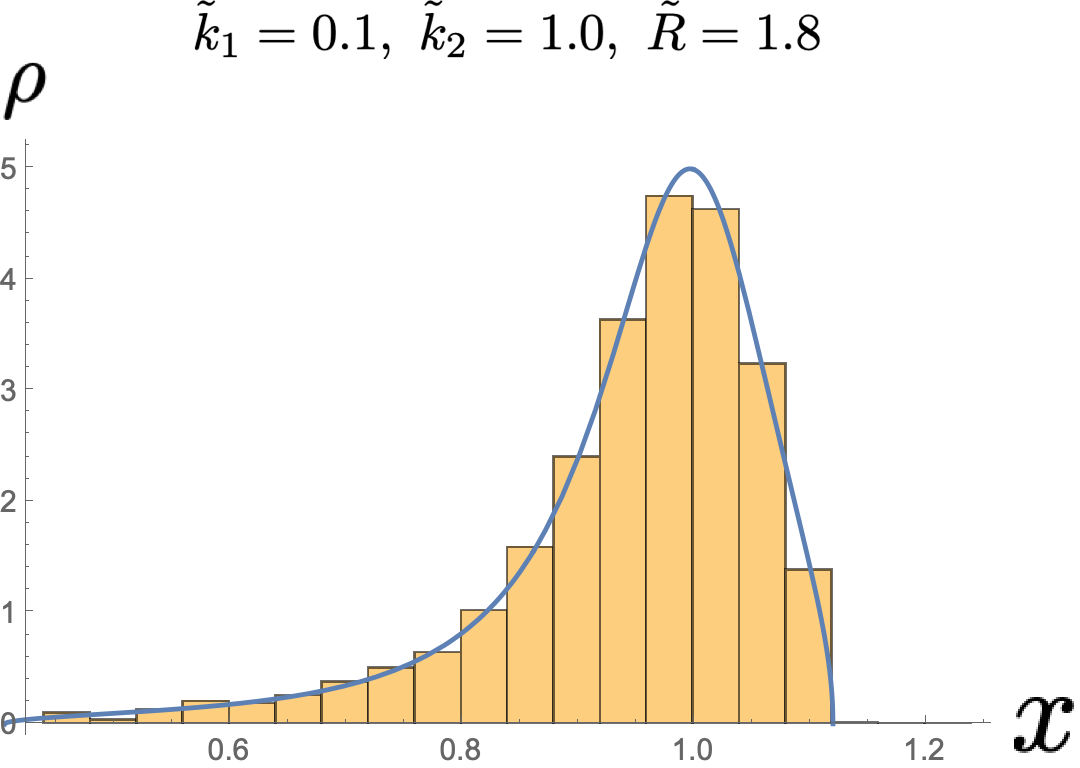}
\includegraphics[width=4cm]{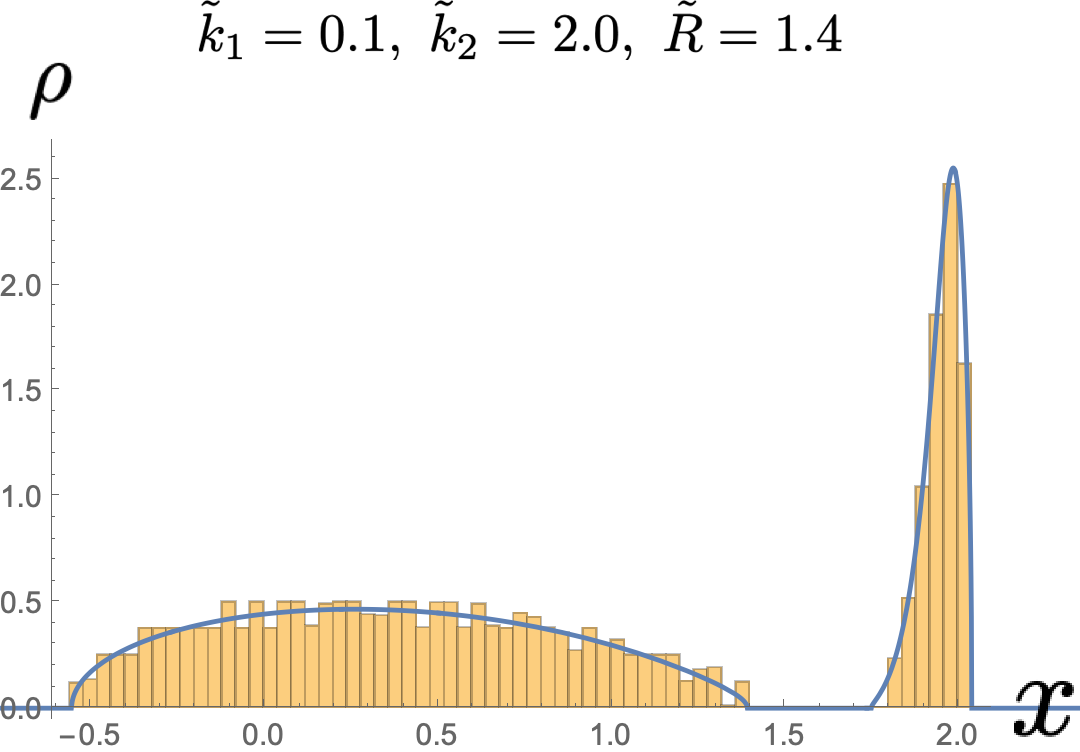}
\includegraphics[width=4cm]{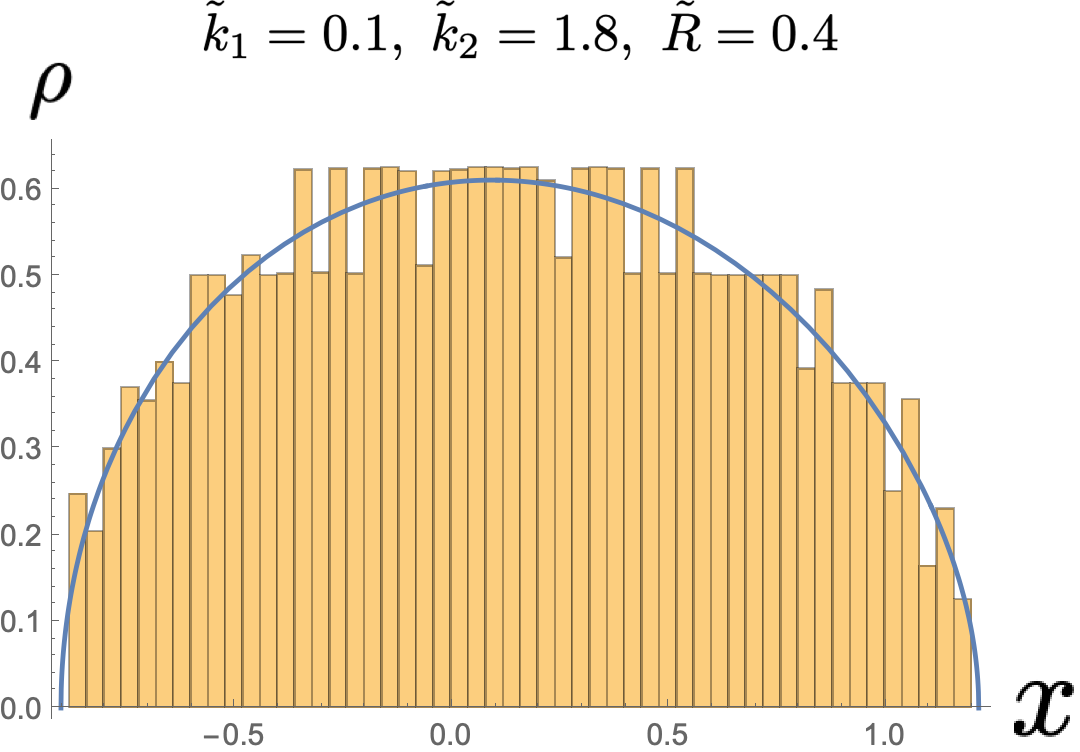}
\caption{The examples of the eigenvalue densities corresponding to the four cases (I), (II), (III), (IV), which
are ordered from the left to the right. 
The results of the Monte Carlo simulations of the system \eq{eq:scoulomb} with $N=200$ 
are shown by the histograms of $\lambda_a$ after the rescaling \eq{eq:rescaling}.
The eigenvalue densities $\rho(x)$ obtained by solving the equations in Section~\ref{sec:sd}
are shown by the solid lines. We take $\alpha=1$.
There are no clear qualitative differences between the cases (I) and (II) of the examples.}
\label{fig:eigenex}
\end{center}
\end{figure}

It seems difficult to obtain explicit solutions to the equations obtained in Section~\ref{sec:sd}. 
In particular, \eq{eq:chemical} is an integral equation and would probably not be explicitly solvable. 
Yet, it is possible to numerically find the solutions. 
For demonstrations, Figure~\ref{fig:eigenex} shows some eigenvalue densities $\rho(x)$
which have been obtained by numerically solving the equations in Section~\ref{sec:sd} for
the four cases in Section~\ref{sec:coulomb}.

These are consistent with the results obtained from the Monte Carlo simulations of the 
aligned Coulomb gas system with $N=200$ in Section~\ref{sec:coulomb}, 
as shown by the histograms of $\lambda_a$ after the rescaling \eq{eq:rescaling}.
We can see that there are no clear qualitative differences between the profiles of (I) and (II), and therefore 
the classifications into (I) or (II) are rather arbitrary in the examples.

\begin{figure}
\begin{center}
\includegraphics[width=10cm]{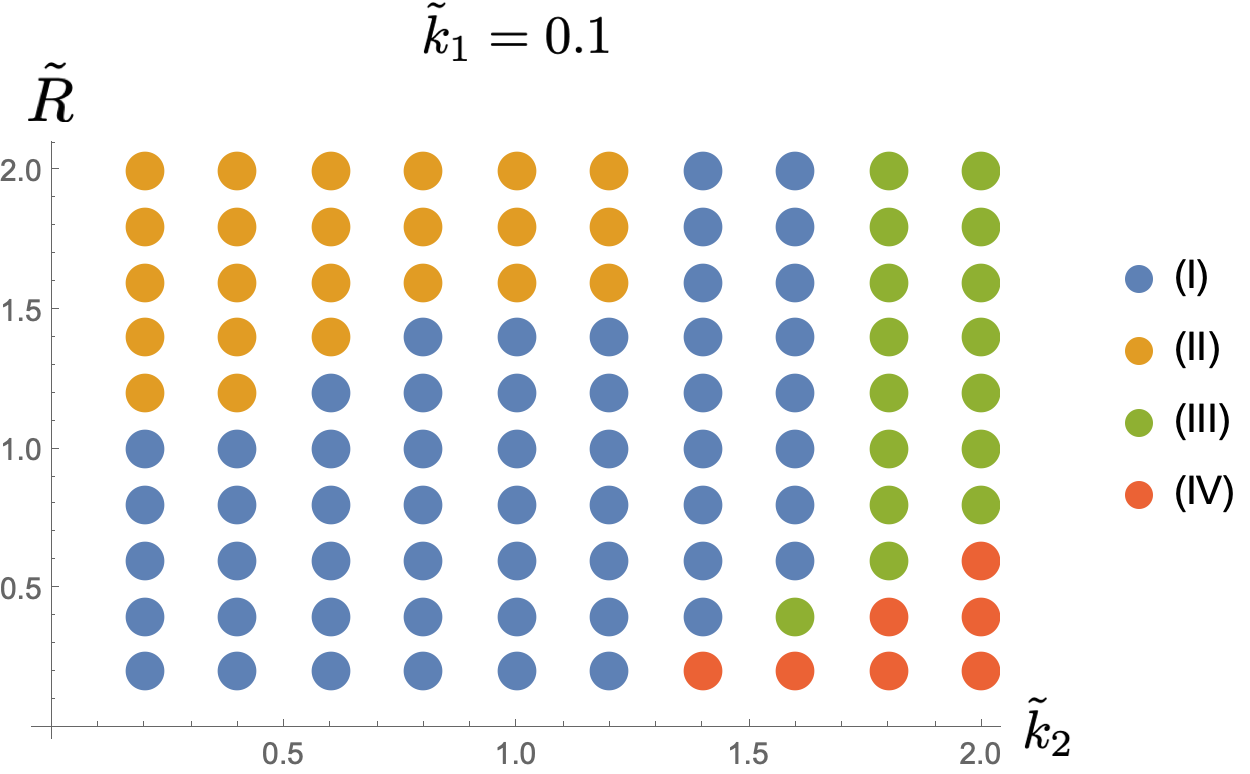}
\caption{The classification of the data points 
into the cases (I)-(IV) by the histograms of $\lambda_a$ for $\tk_1=0.1$ and $\alpha=1$.
The points are taken with interval $0.2$ in the square region, $0.2 \leq \tk_2 \leq 2.0$ and  
$0.2 \leq \tR \leq 2.0$.
The histograms of $\lambda_a$ are obtained 
from the Monte Carlo simulation of the Coulomb gas system for each point of the parameters with $N=200$.
A caution is that the boundary between (I) and (II) is rather arbitrary, because of the lack
of a clear distinction between the two profiles.
}
\label{fig:phasestr}
\end{center}
\end{figure}

In Figure~\ref{fig:phasestr}, the phase structure is shown in the plane of $(\tk_2,\tR)$ 
for $\tk_1=0.1$ and $\alpha=1$.
This is obtained by classifying the histograms of $\lambda_a$ obtained from the Monte Carlo simulations 
of the system \eq{eq:scoulomb} with $N=200$. 
We see that there indeed exist the four cases discussed in Section~\ref{sec:coulomb}.
A caution in this figure is that the boundary between the two cases (I) and (II) is rather arbitrary,
since the profiles of (I) and (II) cannot clearly be distinguished.

Although the distinction between the cases (I) and (II) is not clear in the examples above, 
one can find a reason for separating the two cases in the limit $\tk_1\rightarrow +0$.
To see this, let us consider the explicitly solvable case of $\tk_2=0$.
This case corresponds to the one-cut solution, and the conditions \eq{eq:cond1} can straightforwardly be solved.
The result of $\rho(x)$ is given by
\[
&\rho(x)=\frac{1}{\pi}\left(\frac{\tR\, \tk_1}{(x^2+\tk_1^2)\sqrt{\tk_1^2+a^2}}+2 \alpha\right) \sqrt{a^2-x^2},
\] where $a$ corresponds to $c_+$, and is determined by the equation,
\[
&\tR-\frac{\tR \tk_1}{\sqrt{\tk_1^2+a^2}}+\alpha \, a^2 =1.
\label{eq:conda}
\]
Since the lefthand side is a monotonically increasing function of $a$ with 0 at $a=0$ and 
$+\infty$ at $a=+\infty$, there always exists a unique solution of $a>0$ to the equation. 
One can also show that, by taking the partial derivatives of the lefthand side with respect to the variables, 
there exists no singular behavior of $a$ as a function of $\tk_1$ and $\tR$. 
This is consistent with the former statement above that 
the cases (I) and (II) cannot absolutely be distinguished. 
However, let us take the $\tk_1\rightarrow +0$ limit. In this case, the solution to \eq{eq:conda} 
is explicitly given by 
\[
a=
\left\{ 
\begin{array}{cl}
\sqrt{(1-\tR)/\alpha}, &\tR<1, \\
0,& \tR \geq 1. 
\end{array}
\right. 
\]
A singularity appears at $\tR=1$.
Hence, after taking the limit $\tk_1\rightarrow +0$, (I) and (II) 
can clearly be distinguished by the order parameter $a$.  
In the next section, we will more thoroughly study the limit $\tk_1\rightarrow +0$.

\section{The $\tk_1\rightarrow +0$ limit}
\label{sec:k1zero}
While it does not seem possible to obtain explicit solutions to the equations in Section~\ref{sec:sd}
for general values of the parameters, 
we can obtain explicit solutions by taking the $\tk_1\rightarrow +0$ limit.
Fortunately, this limit is exactly consistent with our main purpose, 
namely, studying the case corresponding to the positive cosmological constant case of CTM, 
as explained in Section~\ref{sec:connection}.
We obtain explicit solutions for for the cases (I), (II), and (III). 
The case (IV) disappears in this limit, because the potential generated by the $-R$ charge
becomes infinitely deep, and there always exist a bunch of eigenvalues around it.

Let us first consider the case (I). 
We assume $\tR<1$ and $c_-<\tk_2<c_+$ for the one-cut solution.
The last assumption and $\tk_1 \rightarrow +0$ leads to 
$\sqrt{\tk_2+i\tk_1-c_+}\,\sqrt{\tk_2+i\tk_1-c_-}=i \sqrt{c_+-\tk_2}\,\sqrt{\tk_2-c_-}$ 
in the first equation of \eq{eq:cond1}, and therefore $c$ must be pure imaginary.
Then the second equation leads to $c_++c_-=0$ in the limit.
Finally, with the third equation, we obtain
\s[
&c=-\frac{i \tR}{2 \sqrt{c_+-\tk_2}\,\sqrt{\tk_2-c_-}},\\
&c_\pm=\pm \sqrt{\frac{1-\tR}{ \alpha}}
\label{eq:sol1}
\s]  
in the limit. 
Note that, because of the assumption, $c_-<\tk_2 < c_+$, this solution is consistent only if 
\[
1-\tR-\alpha \tk_2^2>0.
\label{eq:range1}
\]
By putting the solution \eq{eq:sol1} to \eq{eq:rho1}, $\rho(x)$ 
is obtained as 
\s[
\rho_{\rm (I)}(x)&=\lim_{\tilde k_1\rightarrow +0} \frac{1}{\pi}\frac{\tR\,\tk_1}{ (x-\tk_2)^2+\tk_1^2}+\frac{2 \alpha}{\pi} \sqrt{(1-\tR)/\alpha-x^2}\\
&= \tR\, \delta(x-\tk_2)+\frac{2 \alpha}{\pi} \sqrt{(1-\tR)/\alpha-x^2},
\label{eq:rho1limit}
\s] 
where the domain of $x$ is restricted to the 
positive region of the square root.  One can check $\int_\mathbb{R} dx \rho_{\rm (I)}(x)=1$ holds.

The result \eq{eq:rho1limit} can intuitively be understood by the aligned Coulomb gas picture 
discussed in Section~\ref{sec:coulomb}.
In the $\tk_1\rightarrow +0$ limit, the potential generated by the charge $-R$ at $x=\tk_2$  
becomes infinitely deep, and some of the particles of unit charges are trapped at the location $x=\tk_2$ 
in the limit. 
When $\tR<1$, namely, $N>R$, the number of the particles which are trapped is equal to $R$,
since the particles are trapped until the $-R$ charge is totally screened. 
This concentration of the particles is represented by the first term of \eq{eq:rho1limit}.
Because of the total screening, the $-R$ charge is totally hidden from the remaining $N-R$ particles of unit 
charges, and therefore they form the semi-circle distribution because of the harmonic potential,
that is represented by the second term of \eq{eq:rho1limit}.

Let us next consider $\tR>1$. In this case, since $N<R$, the $-R$ charge at $x=\tk_2$
can only be partially screened. This means that all the particles of unit charges are concentrated 
around $x=\tk_2$ in the limit $\tk_1\rightarrow +0$, which corresponds to the case (II) of Section~\ref{sec:coulomb}.
Therefore, we may naturally assume the behavior, 
\s[ 
 &c_+\sim \tk_2 +b_+ \tk_1,\\
 &c_-\sim  \tk_2- b_- \tk_1,
 \label{eq:assb}
\s]
in the $\tk_1\rightarrow +0$ limit, where $b_-,b_+>0$.
Putting the assumption \eq{eq:assb} into the first equation of \eq{eq:cond1}, we obtain
\[
c\sim -\frac{i \tR}{2 \tk_1 \sqrt{b_+-i}\sqrt{b_-+i}}
\label{eq:asymc}
\]
as the behavior in $\tk_1\rightarrow +0$. 
By putting the assumption \eq{eq:assb} to the second equation of \eq{eq:cond1}, we see that
the real part of $c$ does not diverge in the limit:
\[
c+c^*\sim 2 \alpha \tk_2.
\]
Therefore, this requires $b_+=b_-$ in \eq{eq:asymc}. 
By noting that $(c+c^*)(c_++c_--2 \tk_2)$
has lower order than $\tk_1$, the third equation of \eq{eq:cond1} leads to
\[
b_\pm=\frac{\sqrt{2\tR-1}}{\tR-1},
\]
where we have introduced a notation, $b_\pm:=b_+=b_-$.
By putting the results into \eq{eq:rho1}, we obtain
\[
\rho_{\rm (II)}(x)=\lim_{\tk_1\rightarrow +\infty} 
\frac{(\tR-1) \sqrt{(b_\pm \tk_1)^2-(x-\tk_2)^2}}{\pi ((x-\tk_2)^2+\tk_1^2)}
=\delta(x-\tk_2),
\]
where the domain of $x$ in the middle expression is the positive region of the square root.

Let us next consider the two-cut solution in the $\tk_1\rightarrow +0$ limit. Since the potential generated by
the $-R$ charge becomes infinitely deep, the case (III) with $N>R$ ($\tR<1$) is the only possibility,
and the phase (IV) does not appear.
In the similar spirit as above, it is natural to assume the behavior in  $\tk_1\rightarrow +0$ as
\s[
&d_1\sim-b_1,\\
&d_2\sim -b_2,\\
&d_3\sim -b_0 \tk_1,\\
&d_4\sim b_0 \tk_1, \\
&d\sim -b,
\label{eq:dbrel}
\s]
where $b,b_i$ are all positive. Here, one may start with assuming different 
proportional coefficients for $d_3$ and $d_4$, but they turn out to be the same by the equations,
as we encountered in the previous case.
The ordering of $d$'s restricts $0<b<b_2<b_1$. By putting the
assumptions into \eq{eq:cond2}, we obtain
\s[
&\tR = 2 \alpha b \sqrt{b_1 b_2 (1+b_0^2)}, \\
&\tk_2-b-(b_1+b_2)/2= 0,\\
&\tR+\alpha \left( -b (b_1+b_2)+(b_1-b_2)^2/4 \right)= 1,
\s] 
in the limit. Solving these equations, we obtain
\s[
&b_0=\sqrt{\tR^2/(4 \alpha^2 b^2 b_1 b_2) -1},\\
&b_{1}=\tk_2 -b + \sqrt{(1-\tR)/\alpha+ 2 b (\tk_2-b)},\\
&b_{2}=\tk_2 -b - \sqrt{(1-\tR)/\alpha+ 2 b (\tk_2-b)}.
\label{eq:solb}
\s]

As discussed below \eq{eq:cond2}, the solution \eq{eq:solb} has one free parameter,
which may be chosen to be $b$. It is not totally free and is restricted to a range.  
One condition comes from the ordering $0<b<b_2$ mentioned above, that leads to 
\[
0<b<\frac{1}{2}\tk_2 -\sqrt{\frac{1}{6}\left(\frac{1}{2}\tk_2^2+ \frac{1-\tR}{\alpha}\right)}.
\label{eq:rangeb}
\] 
Note that the presence of this region requires
\[
\alpha \tk_2^2-1+\tR>0.
\label{eq:regioniii}
\]

Another condition comes from $b_0$ to be real in \eq{eq:solb}, that requires $\tR^2>4 \alpha^2 b^2 b_1b_2$. 
To see what this inequality requires, let us introduce 
a function of $b$, $g(b):=4 \alpha^2 b^2 b_1 b_2$, where $b_1,b_2$ are 
given by the functions of $b$ in \eq{eq:solb}.
Then we find its derivative to be 
\s[
g'(b)=4 \alpha^2 b\left(12 (b-\tk_2/2)^2-\tk_2^2-2 (1-\tR)/\alpha\right) >0,
\s]  
where the positivity comes from \eq{eq:rangeb}. 
Since $g(0)=0$, the condition $\tR^2>4 \alpha^2 b^2 b_1b_2$
may also give an additional bound of the form $0<b<b_{max}$ with $g(b_{max})=\tR^2$.
Though the bound is not explicitly determined,
an important matter here is that there exists a finite range of $b$ for a solution, 
if the condition \eq{eq:regioniii} is satisfied. 

By putting \eq{eq:dbrel} and \eq{eq:solb} to \eq{eq:rho2}, and taking the $\tk_1\rightarrow +0$ limit, 
one obtains 
\s[
\rho_{\rm (III)}(x)=\frac{ 2 \alpha (x+b-\tk_2) \sqrt{(x+b_1-\tk_2)(-x-b_2+\tk_2)}}{\pi (x-\tk_2) }
+\tR\left(1-\frac{1}{\sqrt{1+b_0^2}}\right)\delta(x-\tk_2),
\label{eq:rho3}
\s]
where the domain of $x$ for the first term is restricted to the positive region of the square root.
One can check the normalization, $\int dx \rho_{\rm (III)}(x)=1$.

We can also take into account the condition \eq{eq:chemical} for the balance of the chemical potential
between the two bunches. In the $\tk_1\rightarrow +0$ limit, the condition gives
\s[ 
d&=\frac{\int_{d_2}^{d_3} dy\, \frac{y}{y^2+\tk_1^2} \sqrt{y-d_1} \sqrt{y-d_2} \sqrt{d_3-y} \sqrt{d_4-y}}{
\int_{d_2}^{d_3} dy\, \frac{1}{y^2+\tk_1^2} \sqrt{y-d_1} \sqrt{y-d_2} \sqrt{d_3-y} \sqrt{d_4-y}}
\\
&\sim \frac{\int_{d_2}^{d_3} dy\, \sqrt{y-d_1} \sqrt{y-d_2} }{
\int_{d_2}^{d_3} dy\, \frac{1}{y} \sqrt{y-d_1} \sqrt{y-d_2}}\rightarrow 0,
\s]
because $d_3,d_4\rightarrow 0$, where we assume $d_2=-b_2$ is finitely under zero in the limit,
as we will see its consistency below.
By setting $b=-d\rightarrow +0$ in \eq{eq:solb}, we obtain
\s[
&b_0\rightarrow +\infty,\\
&b_{1}=\tk_2 + \sqrt{(1-\tR)/\alpha},\\
&b_{2}=\tk_2  - \sqrt{(1-\tR)/\alpha}.
\label{eq:solb0}
\s]
For this solution, the eigenvalue distribution is given by
\[
\rho_{\rm (III)}^{balance}(x)=\frac{ 2 \alpha \sqrt{(1-\tR)/\alpha-x^2}}{\pi} +\tR\, \delta(x-\tk_2),
\label{eq:rho3b0}
\]
where the domain of $x$ for the first term is restricted to the positive region of the square root.
We again obtain the distribution of the semi-circle law and the concentration at $x=\tk_2$.

\begin{figure}
\begin{center}
\includegraphics[width=7cm]{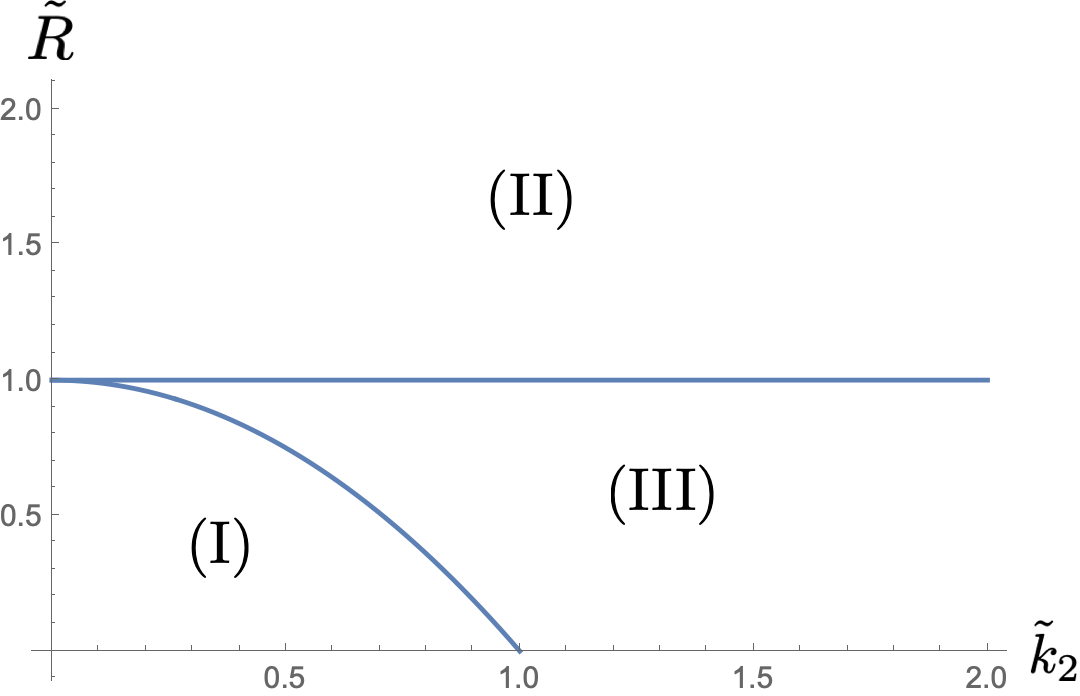}
\caption{The phase structure in the $\tk_1\rightarrow +0$ limit. $\alpha=1$ is taken for the figure.}
\label{fig:phasek1zero}
\end{center}
\end{figure}

The condition \eq{eq:regioniii} for the presence of a solution for (III) is complement to \eq{eq:range1}
for (I).  Therefore, there is a transition line,
\[
\alpha \tk_2^2-1+\tR=0,
\]
between the two phases (I) and (III).
Collecting the results in this section, 
the phase structure in the $\tk_1\rightarrow +0$ limit is given as in Figure~\ref{fig:phasek1zero}.

\section{Implications to the canonical tensor model}

The wave function \cite{Sasakura:2011sq,Sasakura:2012fb} of the canonical tensor model (CTM) 
\cite{Sasakura:2011sq,Sasakura:2012fb} has the interesting property that the peaks of the wave 
function are located at the configurations (the values of the tensor $P_{abc}$) which are 
invariant under Lie group symmetries \cite{Obster:2017pdq,Obster:2017dhx}.
This property is particularly interesting from the perspective of spacetime emergence in CTM,  because this property would provide a natural mechanism for spacetime emergence 
by generating spacetimes as representation spaces of Lie groups.
However, this property has not been fully studied because of some difficulties, 
as explained in Section~\ref{sec:introduction} and Section~\ref{sec:connection}. 
This gives the main motivation for studying the particular simplified matrix model \eq{eq:partition},
which is derived from a matrix version of the wave function of CTM.

The relation between peaks and symmetries is simple in the matrix model \eq{eq:partition}.
The real symmetric $M_{ab}$ becomes Lie-group symmetric, 
when and only when $n$ of the eigenvalues take the same values, 
in which $SO(n)$ is the Lie-group symmetry\footnote{We may also consider the possibilities of
concentrations to multiple values, but this does not occur in the present matrix model, in which
only a concentration at $x=\tk_2$ can occur.}. 
In the aligned Coulomb gas picture of Section~\ref{sec:coulomb},
this corresponds to that $n$ of the particles of unit charges concentrate on a location. 
In fact, in the $\tk_1\rightarrow +0$ limit, $R$ (or $N$ if $N<R$) of the particles concentrate at $x=\tk_2$
to screen the $-R$ charge there, as has explicitly been derived in Section~\ref{sec:sd}.
Therefore, we see the emergence of $SO(R)$ (or $SO(N)$ if $N<R$) symmetry in the matrix model
in the $\tk_1\rightarrow +0$ limit.

The above discussions do not depend on the value of $\tk_2$. However, we have two different 
phases (I) and (III) on the same value of $\tR<1$ as shown in Figure~\ref{fig:phasek1zero}.
In fact there is a qualitative difference between (I) and (III) concerning
the symmetry emergence.

\begin{figure}
\begin{center}
\includegraphics[width=7cm]{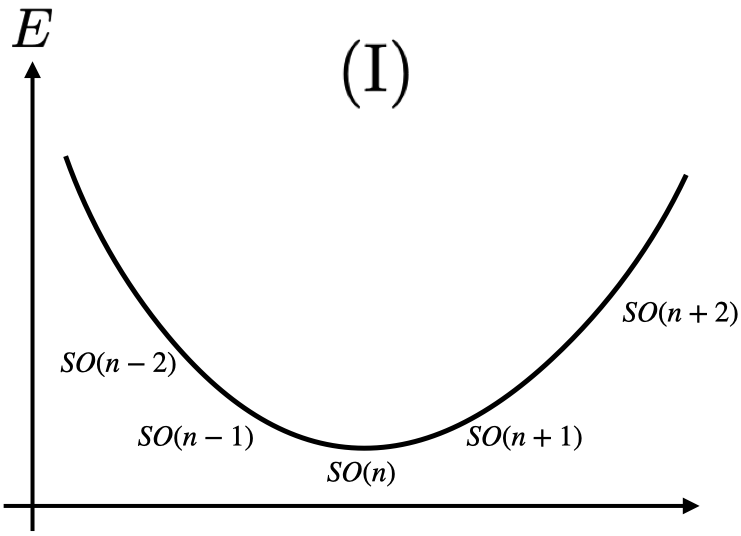}
\hfil
\includegraphics[width=7cm]{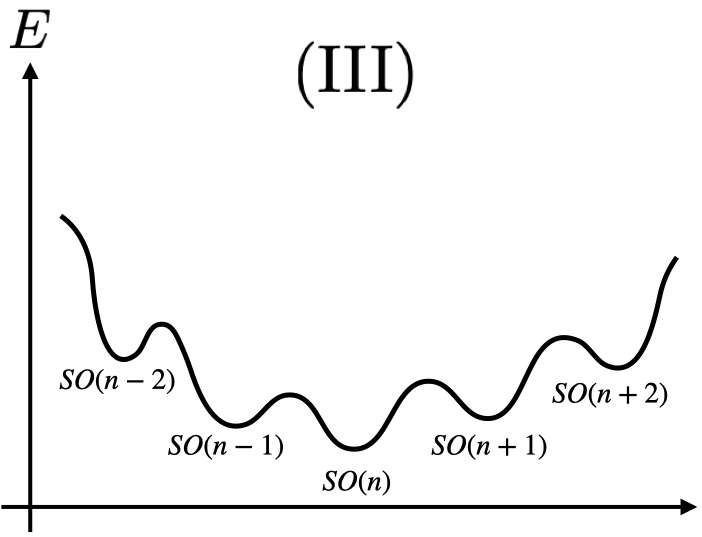}
\caption{An illustration of the difference between the phases (I) and (III) concerning the symmetry.
In phase (I), the symmetry associated to an eigenvalue distribution is ambiguous and 
is subject to changes under perturbations. 
In phase (III), symmetries are definite and stable against perturbations.
In (III) there are metastable states with other symmetries.}
\label{fig:external}
\end{center}
\end{figure}

To see this in the following discussions, 
let us assume that $\tk_1$ is very small $\tk_1 \ll 1$, but the limit $\tk_1\rightarrow +0$ is not 
strictly taken. 
It would still be meaningful to discuss the symmetry enhancement above in an approximate sense,
though the concentration of the eigenvalues is not exactly on $\tilde k_2$ anymore.
Let us first note that, in phase (I), all the eigenvalues are continuously distributed, and  
there are no clear boundary between the eigenvalues near $\tk_2$ and those which are not.
Therefore  the number $n$ of the eigenvalues near $\tk_2$ is ambiguous, meaning that the enhanced 
symmetry is ambiguous in this phase.
Moreover, by adding perturbations to the system,
it would be possible to smoothly move some of the eigenvalues toward or away from $\tk_2$.
Therefore, the symmetry can easily be changed under perturbations.
The situation is illustrated in the left figure of Figure~\ref{fig:external}. 
 
\begin{figure}
\begin{center}
\includegraphics[width=5cm]{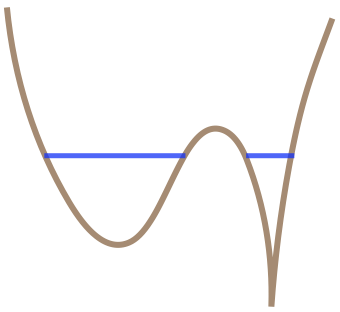}
\caption{The illustration of the potential and the chemical potential corresponding to the case (III).
There is a potential barrier between the two bunches of eigenvalues. The chemical potentials 
take the same value in the figure. If some of the eigenvalues are moved between the bunches, the
chemical potentials get unbalanced, corresponding to metastable states.}
\label{fig:divide}
\end{center}
\end{figure}

On the other hand, in phase (III), the bunch of the eigenvalues around $\tk_2$ is separated from the other 
bunch of the eigenvalues.
In fact, there exists a potential barrier (infinite in the large $N$ limit) for eigenvalues to move between 
the two bunches (the necessary force is proportional to \eq{eq:force}. See Figure~\ref{fig:divide}
for illustration.).
Therefore, in phase (III), the symmetry associated to an eigenvalue distribution is definite and is stable
under perturbations. There are also metastable states which correspond to distributions obtained 
by moving some of the eigenvalues between the bunches.
Other symmetries are associated to the metastable states, since the numbers of the eigenvalues 
in the bunch around $\tk_2$ are different. 
The situation is illustrated in the right figure of Figure~\ref{fig:external}. 

Let us now move on to the discussions on the emergent spaces associated to the symmetries. 
To see this, let us go back to the origin \eq{eq:wavefn} of the matrix model. 
It is clear from the expression that the distribution of $\phi_a$ extends to the directions
of the eigenvectors of $M_{ab}$ whose eigenvalues are near $k_2$.
In other words, when $M_{ab}$ has $n$ eigenvalues near $k_2$,
the distribution of $\phi_a$ forms an $n$-dimensional ball $B^n$ with the radius of 
order $1/\sqrt{k_1}$. The differences between the other eigenvalues and $k_2$ determine
the transverse sizes (the thickness) of the ball.   
The ball gives the representation space of the $SO(n)$ symmetry.
 
\begin{figure}
\begin{center}
\includegraphics[width=7cm]{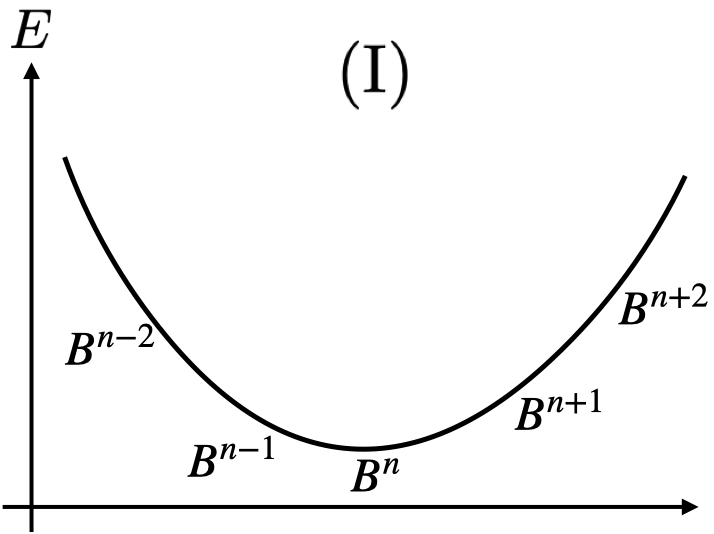}
\hfil
\includegraphics[width=7cm]{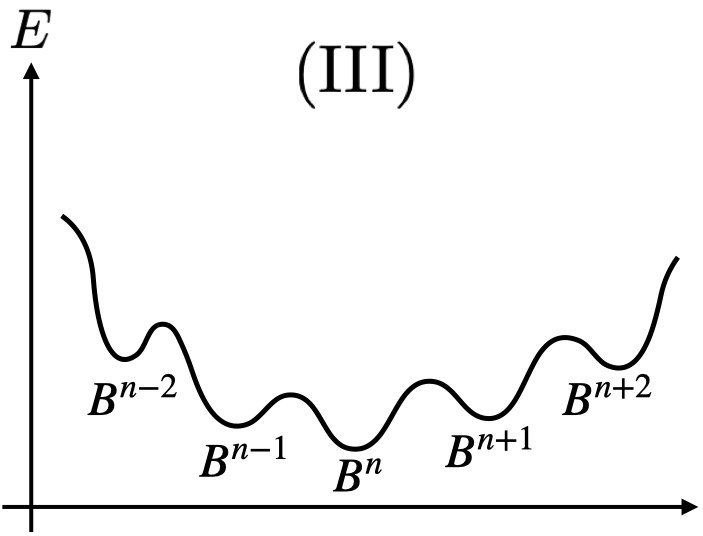}
\caption{An illustration of the difference between the phase (I) and (III) concerning the 
dimension of an emergent space, an $n$-dimensional ball $B^n$.
In phase (I), the dimension is ambiguous and is subject to changes by perturbations. 
In phase (III), symmetries are definite and stable against perturbations.
There are metastable states representing other dimensional balls 
in phase (III).}
\label{fig:ball}
\end{center}
\end{figure}

Again we encounter an ambiguity in phase (I). The sizes of some of the transverse
directions (the thickness) of the ball $B^n$ are actually similar to that of $B^n$, and therefore
the dimension of the ball is ambiguous. In addition, the dimension is subject to changes by perturbations.
On the other hand, in phase (III), the dimension is well determined, because the clear distinction
between the eigenvalues of $M_{ab}$ which are near $k_2$ and those which are not 
provides a hierarchy of the sizes between the ball $B^n$ and its transverse directions. 
The dimension is also stable against perturbations.  
The situation is described in Figure~\ref{fig:ball}.  

Another major outcome from the matrix model is that this gives a few
indirect supports to the previous Monte Carlo results for CTM \cite{Sasakura:2019hql,Obster:2020vfo,Sasakura:2020jis}. 
Here, as explained in Section~\ref{sec:introduction} and Section~\ref{sec:connection}, 
the previous numerical results are for the negative cosmological constant,
corresponding to $k_2=0$ in the matrix model. An important previous result is that 
we have found a continuous phase transition point near\footnote{There is a difference 
of the definition of $R$ by a factor of 2 between the present and previous papers
for the convenience of each.
$2R$ in this paper corresponds to $R$ in the previous papers \cite{Lionni:2019rty,Sasakura:2019hql,Obster:2020vfo,Sasakura:2020jis}.}  
$2R\sim N^2/2$.
This was observed for large $N$ and small $k$, which corresponds to $k_1$ in the matrix model.  
This transition point seems to correspond to the singular point of the matrix model found at $R=N$ in the limit 
$\tk_1=k_1/N\rightarrow +0$, as discussed in the final part of Section~\ref{sec:sol}.
If we identify them as similar points, the phase diagram of the matrix model shown in Figure~\ref{fig:phasek1zero}
suggests that the transition point found for CTM in the previous Monte Carlo studies is 
a common endpoint of phase transition lines which extend into the parameter region of positive 
cosmological constants. This gives a strong motivation for the future study of CTM for the positive 
cosmological constant case, because the transition lines in the matrix model 
are the boundary of the phase (III), which is important for stable symmetries/dimensions of emergent
spaces.

\begin{figure}
\begin{center}
\includegraphics[width=7cm]{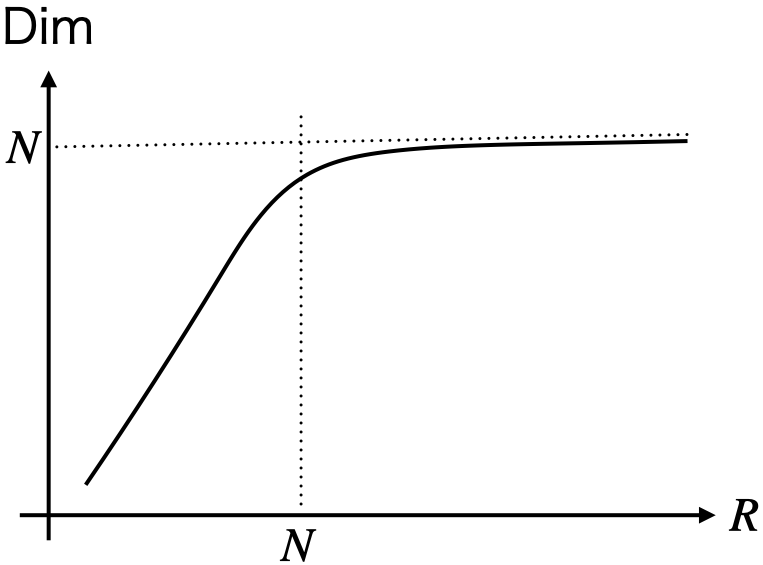}
\hfil
\includegraphics[width=7cm]{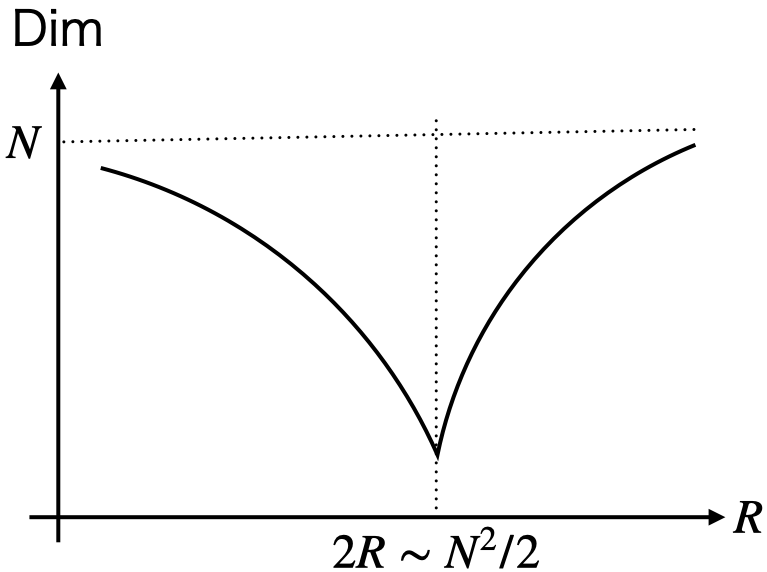}
\caption{The illustrations of the $R$-dependences of the dimensions of $\phi_a$ for the matrix (left) 
and the CTM (right) cases. The right behavior has been obtained in \cite{Obster:2020vfo}. }
\label{fig:dimdep}
\end{center}
\end{figure}

The $R$-dependence shown above of the symmetries and dimensions of the configurations of $\phi_a$
in the matrix model seems in parallel with the $R$-dependence of them observed for CTM
in the previous papers \cite{Sasakura:2019hql,Obster:2020vfo,Sasakura:2020jis}.   
However, there exists a crucial difference. In the matrix model, 
the dimension increases with $R$, reaches its maximum around $R=N$
(see the left figure of Figure~\ref{fig:dimdep}).
On the other hand, in CTM, the dimension takes the minimum at the critical point
(see the right figure of Figure~\ref{fig:dimdep}).  
Currently, there is no understanding for this difference.

\section{Summary and future prospects} 
In this paper, I have studied a matrix model, which is derived from a matrix analogue to the wave function of
the canonical tensor model (CTM), and have shown that the positivity of the cosmological constant is vital 
for the presence of emergent spaces with stable symmetries and dimensions.
  
More precisely, I have studied a one-matrix model of a real symmetric matrix 
with a potential which is a sum of two logarithmic functions and a quadratic one.
This is a simplified toy model obtained by replacing the tensor argument of the wave 
function \cite{Narain:2014cya} of CTM \cite{Sasakura:2011sq,Sasakura:2012fb}
with a matrix and considering the absolute square norm of the wave function. 
The main purpose of the present study is to foresee the possible outcomes of CTM 
for the positive cosmological constant case by studying the toy matrix model for the corresponding case.
The properties of CTM for the positive cosmological constant case have not been studied 
in the previous numerical works \cite{Sasakura:2019hql,Obster:2020vfo,Sasakura:2020jis}
because of some technical difficulties: 
The quantity we need to compute is oscillatory for the positive cosmological constant case, 
suffering from the sign problem of Monte Carlo simulations.
In this paper, it has been shown in the matrix model that the positivity of the cosmological constant is vital 
for the presence of a phase in which symmetries and dimensions of emergent spaces are 
definite and stable.
This result would strongly encourage the future study of the positive cosmological constant case in CTM.

The matrix model has been shown to have a critical point at $R=N$ for $k_2=0$ 
in the limit $\tk_1=k_1/N\rightarrow +0$, as discussed at the end of Section~\ref{sec:sol}.
A similar critical point has been found for CTM at $2R\sim N^2/2$ in the previous works
\cite{Lionni:2019rty,Sasakura:2019hql,Obster:2020vfo,Sasakura:2020jis}.
Therefore the analytical result of the matrix model has given an indirect confirmation to 
the numerical Monte Carlo result in the previous papers 
\cite{Sasakura:2019hql,Obster:2020vfo,Sasakura:2020jis}. 
However, there exists a crucial difference between the two critical points
concerning the dimensional behaviors in $R$, 
as discussed in Section~\ref{sec:connection} (See Figure~\ref{fig:dimdep}). 
This difference should be studied more deeply, since it would 
directly be linked to the mechanism of the emergent phenomena in CTM. 

Though the matrix model presented in this paper does not fully describe the properties of CTM,
it gives intriguing insights for future analysis of CTM.
For instance, the importance of the eigenvalue density profiles for the dynamics of emergent spaces
in the matrix model motivates the study of CTM from the light of the tensor eigenvalue/vector problem \cite{Qi}. 
Various other aspects, such as thermodynamic
properties, of the matrix model remain unexplored, which are also expected to give some insights
into CTM. The matrix model would also provide an arena for developing tools to effectively analyze CTM.
    
\vspace{.3cm}
\section*{Acknowledgements}
The work of N.S. is supported in part by JSPS KAKENHI Grant No.19K03825. 

\appendix
\def\thesection{Appendix \Alph{section}}
\section{A minimal introduction to CTM}
\label{app:ctm}
In this appendix, we will provide minimal information about the canonical tensor model (CTM) for
readers to understand the connection between CTM and the matrix 
model \eq{eq:partition}. 
A more thorough but concise summary of CTM can be found in an appendix of \cite{Sasakura:2019hql}. 

The canonical tensor model (CTM) \cite{Sasakura:2011sq,Sasakura:2012fb} is a tensor model formulated 
as a first-class constrained system 
in the Hamilton formalism. Its dynamical variables are a canonical conjugate pair of real symmetric 
three-index tensors, $Q_{abc}$ and $P_{abc}$\ $(a,b,c=1,2,\ldots,N)$, and 
there are two kinds of first-class constraints ${\cal H}_a$ and ${\cal H}_{ab}$ $(=-{\cal H}_{ba})$,
which form a closed Poisson algebra with dynamical variable dependent structure coefficients.
The canonical quantization of CTM is straightforward \cite{Sasakura:2013wza}, 
and the physical state condition is given by 
$\hat {\cal H}_a|\Psi\rangle=\hat {\cal H}_{ab}|\Psi\rangle=0$,
where $\hat {\cal H}_a$ and $\hat {\cal H}_{ab}$ are the quantized constraints. 
The explicit form of $\hat {\cal H}_a$ is given by
\[
\hat {\cal H}_a=\hat P_{abc} \hat P_{bde}\hat Q_{cde} -\lambda\, \hat Q_{abb}+ i\, \lambda_H \, \hat P_{abb},
\]
where $\lambda$ is identified with the cosmological constant of general relativity (GR) from the 
equivalence between the $N=1$ CTM and the minisuperspace approximation of GR \cite{Sasakura:2014gia}. 
$\lambda_H$ is uniquely determined by the hermiticity of $\hat {\cal H}_a$ as $\lambda_H=(N+2)(N+3)/2$. 
The wave function which represents the exactly solved physical state is given by 
\[
\Psi_{CTM}(P):=\langle P | \Psi\rangle=\varphi_{CTM}(P)^{\lambda_H/2},
\]
where
\[
\varphi_{CTM}(P):=\int_{\cal C} d\tilde \phi  \prod_{a=1}^N d\phi_a \, \exp\left[ 
i P_{abc} \phi_a \phi_b \phi_c -i\,k \,\phi^2 \tilde \phi+i\, \tilde \phi^3/3
\right].
\label{eq:varphictm}
\]
Here $k$ has the same sign as the cosmological constant, because $\lambda\propto k^3$.
The integration contour ${\cal C}$ can be taken in various manners with an infinite extent as far as 
the integration converges.
We consider the naive choice ${\cal C}=\mathbb{R}^{N+1}$
with a regularization, which can for example be taken as the one in \cite{Obster:2017dhx}\footnote{A more
rigorous way of defining the integral would be to express the integration contour 
as a sum of Lefschetz thimbles \cite{Witten:2010cx}.}.
 
\begin{figure}
\begin{center}
\includegraphics[width=7cm]{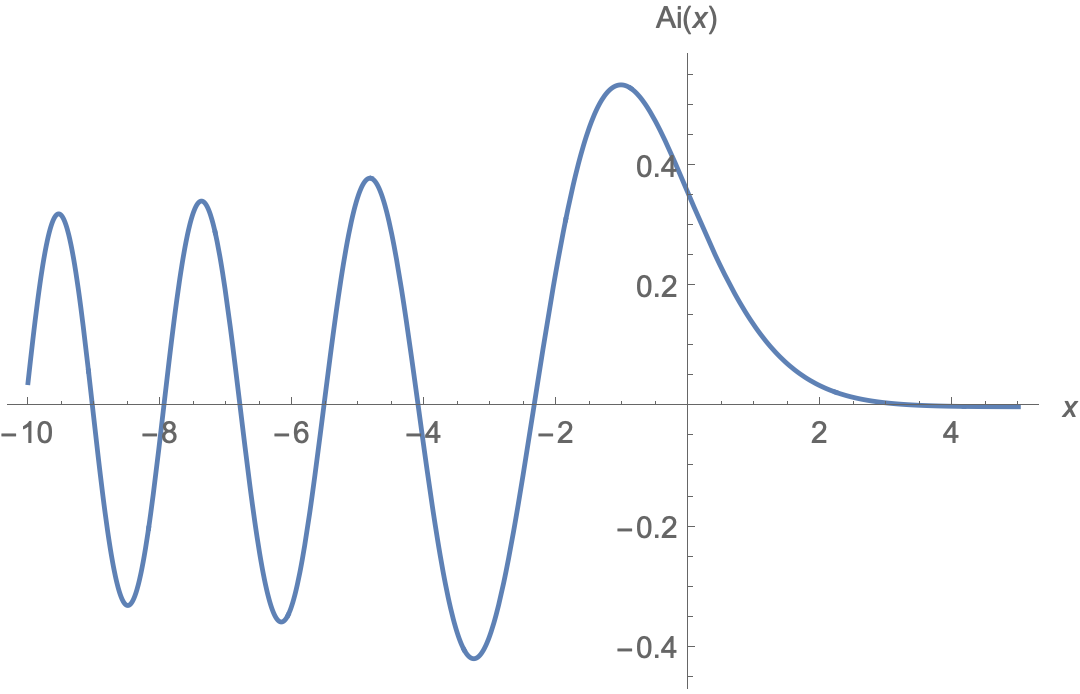}
\caption{The Airy Ai function}
\label{fig:airy}
\end{center}
\end{figure}

The integration over $\tilde \phi$ in \eq{eq:varphictm} provides an expression, 
\[
\varphi_{CTM}(P) = const.  \int_{\mathbb{R}^D} \prod_{a=1}^N d\phi_a \, \exp\left( 
i P_{abc}\phi_a \phi_b \phi_c \right) {\rm Ai}(- k\, \phi^2),
\label{eq:varphiai}
\]   
where ${\rm Ai}(\cdot)$ denotes the Airy Ai function.  For $k>0$ corresponding 
a positive cosmological constant, ${\rm Ai}(- k\, \phi^2)$ is an oscillatory function of $\phi^2$, 
while, for $k<0$ corresponding to the negative cosmological constant, this is a damping function of $\phi^2$
(See Figure~\ref{fig:airy}).  

\end{document}